\documentclass[12pt]{article}
\usepackage{amsmath}
\usepackage{amssymb}
\date{}
\begin{document}

\title{On the Solution of the Van der Pol Equation}

\author{Mayer Humi\\
Department of Mathematical Sciences,\\
Worcester Polytechnic Institute,\\
100 Institute Road,\\
Worcester, MA  0l609}

\maketitle

\begin{abstract}
We linearize and solve the Van der Pol equation (with additional 
nonlinear terms) by the application of a generalized form of Cole-Hopf 
transformation. We classify also Lienard equations with low order polynomial
coefficients which can be linearized by this transformation. 
\end{abstract}

\thispagestyle{empty}

\newpage

\section{Background}

The Van der Pol equation [1] was derived originally to model electrical 
circuits in vacuum tubes. However since then it has been used in both 
the physical [2-6] and biological sciences. For instance, in biology, 
Fitzhugh and Nagumo [7] considered a two dimensional version of this
equation as a model for the action potentials of neurons. 

While there is no existing general theory for integrating nonlinear
ordinary differential equations (ODEs), the derivation of exact 
(closed form) solutions for these equations is a non-trivial and important 
problem. Some methods to this end are 
Lie, and related symmetry methods [8]. Many other ad-hoc methods for solving 
nonlinear ODES and partial differential equations were independently 
suggested during the last decades.
Among these methods Cole-Hopf transformation [9,10,11,12] has been used 
originally to linearize the Burger's equation [11] which has various 
applications in applied mathematics and theoretical physics. 
Recently however, we introduced [13] a generalization of Cole-Hopf 
transformation and used it to linearize and solve various nonlinear ordinary 
differential equations e.g Duffing equation. We showed also that many of the 
special functions of mathematical physics are exact solutions for a class 
of nonlinear equations.

In this paper we apply in Sec. $2$ the same generalization of Cole-Hopf
transformation to linearize the Van der Pol equation with additional quadratic 
and cubic nonlinear terms. These terms can be considered as perturbations
to the original equation. Next in Sec. $3$ we present explicit solution 
(in closed analytic form) to this equation with no external forcing.
Solutions to this equation when external forcing term is included
can be obtained also by this method. In Sec. $4$ we consider the 
linearization of Lienard equations with low order polynomial coefficients
(in the dependent function). 
We end up in Sec. $5$ with some conclusions.

\setcounter{equation}{0} 
\section{The Perturbed Van der Pol Equation}

The original form of Van der Pol equation with no forcing can be written as
\begin{equation}
\label{3.1}
\psi(x)''=\mu(\beta-\psi(x)^2)\psi(x)'-\alpha\psi(x)
\end{equation}
where $\alpha$ and $\beta$ and $\mu$ are arbitrary constants.
In this paper however we consider a more general form of this equation
with additional nonlinear and forcing terms
\begin{equation}
\label{3.2}
\psi(x)''=\mu(\beta-\psi(x)^2)\psi(x)'-\alpha\psi(x)+v(x)\psi(x)^2+h(x)\psi(x)^3+g(x)\psi(x)^4+f(x).
\end{equation}

We shall say that the solutions of (\ref{3.2})
and
\begin{equation}
\label{3.3a}
\phi(x)''=U(x)\phi(x)
\end{equation}
are related by a generalized Cole-Hopf transformation if we can 
find a function $P(x)$ so that 
\begin{equation}
\label{3.4a}
\psi(x)=P(x)+\frac{\phi(x)^{\prime}}{\phi(x)}.
\end{equation}

To classify those nonlinear equations of the form (\ref{3.2}) whose solutions 
can be obtained from those of the linear equation (\ref{3.3a}) we differentiate 
(\ref{3.4a}) twice and in each step replace the second order derivative of 
$\phi(x)$ using (\ref{3.3a}). Substituting these results in (\ref{3.2}) 
leads to the following equation
\begin{eqnarray}
\label{3.3}
a_4(x)\left(\frac{\phi(x)'}{\phi(x)}\right)^{4}+
a_3(x)\left(\frac{\phi(x)'}{\phi(x)}\right)^{3}+
a_2(x)\left(\frac{\phi(x)'}{\phi(x)}\right)^{2}+
a_1(x)\frac{\phi(x)'}{\phi(x)}+a_0(x)=0
\end{eqnarray}
where
\begin{equation}
\label{3.4}
a_4(x)=-\mu-g(x)
\end{equation}
\begin{equation}
\label{3.5}
a_3(x)=-[2\mu+4g(x)]P(x)-h(x)+2
\end{equation}
\begin{equation}
\label{3.6}
a_2(x)=\mu P(x)'-(6g(x)+\mu)P(x)^2-3h(x)P(x)-v(x)+\mu U(x)+\mu\beta
\end{equation}
\begin{equation}
\label{3.7}
a_1(x)=-4g(x)P(x)^3-3h(x)P(x)^2+[2\mu P(x)'-2v(x)+2\mu U(x)]P(x)-2U(x)+\alpha
\end{equation}
\begin{eqnarray}
\label{3.8}
a_0(x)&=&P(x)^{\prime\prime}+\mu[P(x)^2-\beta)P(x)^{\prime}+U(x)^{\prime}-g(x)P(x)^4 \\ \notag
&&-h(x)P(x)^3+[\mu U(x)- v(x)]P(x)^2+\alpha P(x)-\mu\beta U(x)-f(x)
\end{eqnarray}

To satisfy (\ref{3.3}) it is sufficient to let $a_i(x)=0$, $i=1,2,3,4$.
From $a_4=0$ we obtain
\begin{equation}
\label{3.9}
g(x)=-\mu. 
\end{equation}
Using this result and $a_3(x)=0$ we solve for $h(x)$ (in terms of $P(x)$)
\begin{equation}
\label{3.10}
h(x)=2(\mu P(x)+1). 
\end{equation}
The condition $a_2(x)=0$ can be solved then for $v(x)$
\begin{equation}
\label{3.11}
v(x)=\mu P(x)'+\mu U(x)-[\mu P(x)+6]P(x)+\mu\beta.
\end{equation}
substituting (\ref{3.9})-(\ref{3.11}) in $a_1=0$ we solve for $U(x)$ 
in terms of $P(x)$
\begin{equation}
\label{3.12}
U(x)=3P(x)^2-\mu\beta P(x)+\frac{\alpha}{2}
\end{equation}
Finally using the expressions derived in (\ref{3.9})-(\ref{3.12}) in 
$a_0=0$ we have
\begin{equation}
\label{3.13}
f(x)=P(x)^{\prime\prime}-2\mu\beta P(x)^{\prime}+[6P(x)'+\alpha+\mu^2\beta^2]P(x)+4[P(x)-\mu\beta]P(x)^2-\frac{\mu\beta\alpha}{2}
\end{equation}
This equation can be solved (in principle) for $P(x)$ for a given $f(x)$ 
or it can be used to determine $f(x)$ for apriori choice of $P(x)$. 
Eqs. (\ref{3.10})-(\ref{3.12}) can be solved then (in reverse order) 
to compute the functions $U(x),v(x)$ and $h(x)$.

Using this algorithm we provide in the next section explicit solutions to 
(\ref{3.2}) under various conditions.

\setcounter{equation}{0} 
\section{Solutions to the Van der Pol Equation}

\subsection{Van der Pol Equation with No External Forcing}

When $f(x)=0$ the general solution of (\ref{3.13}) for $P(x)$ is
\begin{equation}
\label{4.1}
P(x)=\frac{2C_1\mu\beta e^{\frac{\mu\beta x}{2}}+C_2(\mu\beta+k)e^{\frac{kx}{2}}+
(\mu\beta-k)e^{-\frac{kx}{2}}}
{4[C_1e^{\frac{\mu\beta x}{2}}+C_2e^{\frac{kx}{2}}+e^{-\frac{kx}{2}}]}
\end{equation}
where $k^2=\mu^2\beta^2-4\alpha$ and $C_1,C_2$ are arbitrary constants.
Similar (but more cumbersome) expressions can be obtained for $P(x)$ when
$f(x)=c$ where $c$ is a non zero constant.

It is now only a matter of simple algebra to compute the general form of
the functions $U(x),v(x)$ and $h(x)$ and then derive solutions to (\ref{3.2})
by solving the linear equation (\ref{3.3a}). 

We consider some simple cases.

{\bf Case} 1: $C_1=C_2=0$.
In this case (\ref{4.1}) reduces to 
\begin{equation}
\label{4.2}
P(x)=\frac{\mu\beta-k}{4}.
\end{equation}
The resulting expressions for $U(x),v(x)$ and $h(x)$ are
\begin{equation}
\label{4.3}
U(x)=\alpha/2+\frac{(3k+\mu\beta)(k-\mu\beta)}{16}
\end{equation}
\begin{equation}
\label{4.4}
v(x)=-\frac{\mu^3\beta^2}{8}+\frac{\mu}{2}\left(\alpha-\beta+\frac{k^2}{4}\right)
+\frac{3k}{2}
\end{equation}
\begin{equation}
\label{4.5}
h(x)=\frac{\mu}{2}(\mu\beta-k)+2.
\end{equation}
With these settings the general solution of (\ref{3.3a}) is 
\begin{equation}
\label{4.5a}
\phi(x)=C_3\cos(\omega x)+C_4\sin(\omega x)
\end{equation}
where
$\omega=\frac{1}{4}\sqrt{\mu^2\beta^2+2\mu\beta k-3k^2-8\alpha}$.
These solutions are related to the solutions of (\ref{3.2}) by
the transformation (\ref{3.4a}). 

{\bf Case} 2: Assume $k=0$ and $C_1=0$.

In this case 
\begin{equation}
\label{4.6}
P(x)=\frac{\mu\beta}{4},\,\,\, U(x)=\frac{\alpha}{2}-\frac{\mu^2\beta^2}{16}
,\,\,\,\, h(x)=\frac{\mu^2\beta}{2}+2,\,\,\, 
v(x)=\frac{\mu}{2}\left(\alpha-\beta-\frac{\mu^2\beta^2}{4}\right)
\end{equation}
The solution for $\phi(x)$ is in the same form as in (\ref{4.5a}) with
$\omega=\frac{1}{4}\sqrt{\mu^2\beta^2-8\alpha}$.

{\bf Case} 3: Assume $\alpha=0$.

When $\alpha=0$, $k=\pm \mu\beta$. (In the following we consider only
the plus sign). Under this constraint (\ref{4.1}) reduces to
\begin{equation}
\label{4.7}
P(x)=\frac{(C_1+C_2)\mu\beta e^{\frac{\mu\beta x}{2}}}
{(C_1+C_2+e^{-\mu\beta x}}
\end{equation}
Using (\ref{3.12}) yields
\begin{equation}
\label{4.8}
U(x)=-\frac{c\mu^2\beta^2(e^{-\mu\beta x}-c)}{e^{-\mu\beta x}+c)^2}
\end{equation}
where $c=C_1+C_2$
Similarly (\ref{3.11}) and (\ref{3.12}) lead to
\begin{equation}
\label{4.9}
v(x)=\frac{\mu\beta(e^{-\mu\beta x}-2c)}{e^{-\mu\beta x}+c},\,\,\,\,
h(x)=2+\frac{\mu^2\beta c}{e^{-\mu\beta x}+c}
\end{equation}
With this data the general solution for (\ref{3.3a}) is
\begin{equation}
\label{4.10}
\phi(x)=\frac{C_3+C_4(ce^{\mu\beta x}+\mu\beta x)}{\sqrt{1+ce^{\mu\beta x}}}
\end{equation}

\setcounter{equation}{0}
\subsection{Van der Pol Equation with External forcing}

When the forcing function $f(x)$ is not zero or a constant (\ref{3.13})
can not be solved in closed form for $P(x)$. It is expedient in this 
case to reverse the process and classify those equations of the form 
(\ref{3.2}) which can be linearized for a given $P(x)$.

To make a proper choice of $P(x)$ we observe that in order to obtain
closed form solutions for $\psi(x)$ the function $U(x)$ has to be in a form
which enables the solution of (\ref{3.3a}) to be expressed in "simple form".

To carry this out it is easy to see from (\ref{3.12}) that when 
$$
U(x)= 3g(x)^2-\mu\beta g(x)+\frac{\alpha}{2},
$$
(where g(x) is an arbitrary smooth function) then
$$
P(x)= g(x),\,\,\, or \,\,\, P(x)=-g(x)+\frac{\mu\beta}{3}
$$
The computation of the functions $f(x)$, $v(x)$ and $h(x)$ can be carried out 
then by substitution. We present some examples.

{\bf Example} 1: Let $g(x)=ax$ where a is a constant. We then have
$$
U(x)=3a^2x^2+a\mu\beta x-\frac{\alpha}{2},
$$
$$
f(x)=-4a^3x^3+a\left(6a-\alpha+\frac{\mu^2\beta^2}{3}\right)x-
\frac{\mu\beta\alpha}{6}+\frac{\mu^3\beta^3}{27}
$$
The general solution for $\phi(x)$ can be expressed then in terms of
Hypergeometric functions. 

{\bf Example} 2: let $g(x)=tan(x)$. In this case we obtain an oscillatory 
forcing function and the solutions of (\ref{3.3a}) for $\phi(x)$ can be 
expressed again in terms of Hypegeometric functions.

\section{Polynomial Lienard Equations}

The general form of Lienard equations is
\begin{equation}
\label{5.1}
\frac{d^2\psi}{dt^2}+f(\psi)\frac{dx}{dt}+g(\psi)=0 
\end{equation}
Since Van der Pol equation with no external forcing belongs to this
class of equation it is appropriate to ask if the same method used to solve
(\ref{3.2}) can be applied to this more general class of equations. In this 
section we explore this application to (\ref{5.1}) when $f(x)$ and $g(x)$ are 
respectively a second and fourth order polynomials
$$
f(x)=\displaystyle\sum_{i=0}^2 c_i\psi(x)^i,\,\,\,\,
g(\psi)=\displaystyle\sum_{i=0}^4 b_i\psi(x)^i
$$  
Applying the transformation (\ref{3.4a}) with $\phi(x)$ satisfying 
(\ref{3.3a}) to (\ref{5.1}) leads to an equation similar to (ref{3.3}).
which can be satisfied if we set $a_i=0$ for $i=0,\ldots 4$.
We solve this set of equations for $b_i$ in terms of $a_i$, $P(x)$
and $U(x)$. We obtain the following relations:
\begin{equation}
\label{5.2}
b_4(x)=c_2(x),\,\,\,\, b_3(x)=c_1(x)-2c_2(x)P(x)+2,\,\,\,\,
\end{equation}
\begin{equation}
\label{5.3}
b_2(x)=c_2(x)P(x)^2-2(3-c_1(x))P(x)-c_2(x)(P(x)'-U(x))+c_0(x)
\end{equation}
\begin{equation}
\label{5.4}
b_1(x)=(6+c_1(x))P(x)^2-2c_0(x)P(x)-c_1(x)P(x)'-(c_1(x)+2)U(x),\,\,\,\,
\end{equation}
\begin{equation}
\label{5.5}
b_0(x)=P(x)''-c_0(x)P(x)'+U(x)'-2P(x)^3+2P(x)U(x)+c_0(x)(P(x)^2-U(x))
\end{equation}

For apriori choice of the functions $c_i(x)$ and $U(x),P(x)$ these 
equations can be solved for the $b_i(x)$. When these relations hold 
we obtain an equation of the form (\ref{5.1})
whose solutions are related to those of (\ref{3.3a}) by the 
transformation (\ref{3.4}).

An interesting case arises when $b_0=0$. Under this condition (\ref{5.5})
will be satisfied if
\begin{equation}
\label{5.6}
U(x)=P(x)^2-P(x)'
\end{equation}
which is a Riccati equation and $a_0(x)$ remains a free parameter..

\section{Conclusions}

We demonstrated in this paper that a generalized form of Cole-Hopf 
transformation (\ref{3.4}) can be used to find solutions of the perturbed
Van der Pol equation without forcing. The method can be used also to 
find solutions of this equation with forcing. However in this case one 
can not specify apriori the forcing term.  

\section*{References}
\begin{itemize}

\item[1] B. Van der Pol- A theory of the amplitude of free and forced triode 
vibrations, Radio Rev. 1 (1920) pp 701-710

\item[2] J. Guckenheimer- Dynamics of the Van der Pol Equation,
IEEE Trans. Circuits and Systems, Vol cas-27 pp.983-989 (1980)

\item[3] J. Guckenheimer and P. Holmes-Nonlinear Oscillations, 
Dynamical Systems,and Bifurcations of Vector Fields, 
Springer-Verlag, New York 1990

\item[4] G.M. Moremedi, D.P. Mason, V.M. Gorringe- On the limit cycle of a 
generalized van der Pol equation, International Journal of 
Non-Linear Mechanics, Volume 28, Issue 2, March 1993, Pages 237-25

\item[5] Yajie Li1, Ben T. Nohara, and Shijun Liao- Series solutions of coupled 
Van der Pol equations by means of homotopy analysis method, J. Math. Phys. 
51, 063517 (2010); http://dx.doi.org/10.1063/1.3445770 (12 pages)

\item[6] Raimond A. Struble and John E. Fletcher- General Perturbational 
Solution of the Harmonically Forced van der Pol Equation,
J. Math. Phys. 2, 880 (1961), DOI:10.1063/1.1724236

\item[7] FitzHugh R. (1955) Mathematical models of threshold phenomena in the 
nerve membrane. Bull. Math. Biophysics, 17 pp. 257-278

\item[8] C. Gu, H. Chaohao and Z. Zhou - Darboux Transformations in 
Integrable Systems, Springer, New-York (2005)

\item[9] E. Hopf - The partial differential equation $u_t+uu_x=u_{xx}$,
Commun. Pure Appl. Math {\bf 3}, pp.201-230 (1950)

\item[10] J.D. Cole, On a quasi-linear parabolic equation occurring in 
aerodynamics, Quart. Appl. Math. {\bf 9}, pp.225-236 (1951)

\item[11] P. L. Sachdev-A generalized Cole-Hopf transformation for 
nonlinear parabolic and hyperbolic equations,
ZAMP {\bf 29}, No 6 (1978), pp. 963-970, DOI: 10.1007/BF01590817

\item[12] B. Gaffet-On the integration of the self-similar equations and
the meaning of the Cole-Hopf Transformation J. Math. Phys. 27, 2461 (1986)

\item[13] M. Humi -A Generalized Cole-Hopf Transformation
for Nonlinear ODES, J. Phys. A: Math. Theor. 46 (2013) 325202 (14pp)

\end{itemize}

\end{document}